\documentclass[a4paper,12pt]{article}

\usepackage[left=2.5cm,right=2.5cm,top=2.5cm,bottom=2.5cm]{geometry}
\usepackage{graphicx}
\usepackage{amssymb}
\usepackage{multirow}
\usepackage{amsmath}
\usepackage{psfrag}
\usepackage{setspace}
\usepackage{subcaption}
\usepackage[colorlinks=true,bookmarks=true]{hyperref}
\usepackage{float}
\usepackage{cite}
\usepackage[utf8]{inputenc}

\hypersetup{citecolor=blue}
\newcommand{\dif}{\mathrm{d}}
\newcommand{\Eqref}[1]{(\ref{#1})}
\newcommand{\half}{\frac{1}{2}}
\newcommand{\expo}[1]{\mathrm{e}^{#1}}
\newcommand{\brac}[1]{\left(#1 \right)}
\newcommand{\sbrac}[1]{\left[#1\right]}

\newcommand{\Vcal}{\mathcal{V}}

\newcommand{\Lcal}{\mathcal{L}}
\newcommand{\Ucal}{\mathcal{U}}

\begin{document}

\title{Rotation curves and orbits in the scalar field dark matter halo spacetime}
\author{Yen-Kheng Lim\footnote{E-mail: yenkheng.lim@gmail.com}\\\textit{\normalsize{Department of Mathematics, Xiamen University Malaysia,}}\\\textit{\normalsize{43900 Sepang, Malaysia}}}
\date{\normalsize{\today}}
\maketitle

\begin{abstract}
  The spacetime satisfying the flat curve condition for galaxies is shown to be the zero mass limit of the dilaton blackhole with an exponential potential. We derive the geodesic equations and by studying rotational curves and light deflection, we find that the presence of the dilaton increases the gravitational effects overall but diminishes the contribution from the central mass when the distances are close to the black hole. Also, the dilaton reduces the size of innermost stable circular time-like orbits, while the radius of the photon sphere may be larger or smaller than its Schwrazschild counterpart, depending on the strength of the dilaton field. We also show that a generalisation of the flat-curve-condition spacetime considered elsewhere in the literature does not solve the Einstein equation. 
\end{abstract}

\section{Introduction} \label{intro}

The discrepancy between the observed galactic rotation curves and the expectation based on luminous matter \cite{Rubin:1980zd,Sofue:2000jx} has long been an oustanding problem in gravitational physics. Attempts to resolve this issue can be broadly split into two approaches, namely, (i) to modify the theory of gravity, and (ii) postulating the presence of dark matter in the galactic halo.

In the latter approach, observations lead one to conclude that the dark matter is likely to take a spherical distribution, and takes up a large proportion of the total galactic mass over the luminous matter \cite{Persic:1995ru}. Coupled with the observation that the measured rotational speeds of the stars appear to tend to a constant against distance, this provides some clues to determine the possible type of geometry which should describe spacetime around the galactic halo \cite{CabralRosetti:2001bh,Matos:2001pz,Matos:2000jr,Rahaman:2010xs,Diez-Tejedor:2013sza,Dey:2014gka}. 

In the present work, we will frequently make reference to a particular spacetime proposed by Matos, Guzm\'{a}n, and N\'{u}\~{n}ez (MGN) \cite{Matos:2000ki}, wherein the \emph{flat curve condition} is imposed to fix the form of a static, spherically-symmetric ansatz for the spacetime. The metric obtained by MGN has the form 
\begin{align}
 g_{tt}\propto r^l, \label{IntroEq1}
\end{align}
where $r$ is the radial coordinate and $l$ is a constant that depends on the tangential speed of the flat curve. If the metric were to describe the spacetime of a galaxy with a flat rotation curve, it is required that $l\sim 10^{-6}$ to be consistent with the observed rotational speeds. Having Eq.~\Eqref{IntroEq1}, MG considered a simple scalar field model which would produce such a spacetime, and concluded that the required ingredients are a scalar field dark matter (SFDM) minimally coupled to gravity with a Liouville-type potential. In this solution, the $g_{rr}$ part obtained upon solving the Einstein equations contains an integration constant which was set to zero.

Subsequently, Nandi et al. \cite{Nandi:2009hw} considered a generalisation of the MGN solution where the aforementioned constant is \emph{not} set to zero. With this non-zero integration constant, the $g_{rr}$ is now a non-trivial function. It is then shown that this configuration provides a more realistic energy and pressure density for the dark matter distribution. However, as we will show in the Appendix, this solution cannot be exact, as it does not solve the Einstein's equation exactly. Nevertheless the constant required in \cite{Nandi:2009hw} is small, to the order $\sim10^{-7}$ and hence may perhaps be considered an approximate solution.

An Einstein gravity theory minimally coupled with a scalar field plus a Liouville potential has been well studied in the literature, and many exact solutions have been obtained \cite{Chan:1995fr,Cai:1997ii,Charmousis:2001nq,Dehghani:2004jk,Agop:2005np,Abdolrahimi:2011zg}. A spacetime of relevance to this paper is the dilaton black hole solution given by Chan et al. \cite{Chan:1995fr}, which reduces to the MGN solution in its zero mass limit. Therefore, as an extension of the results of MGN, we shall consider the geodesics of the dilaton black hole spacetime, including both time-like and null particles.

The rest of this paper is organised as follows. In Sec.~\ref{action} we review the action for Einstein-dilaton gravity with a Liouville-type potential along with the dilaton black-hole solution. Then the geodesic equations for this spacetime is derived in Sec.~\ref{geodesics}, followed by Sec.~\ref{curves} where we explore circular orbits and some examples of rotation curves for time-like particles. Null orbits and gravitational lensing is briefly considered in Sec.~\ref{lensing}. Some concluding remarks are given in Sec.~\ref{conclusion}. In Appendix \ref{MGN}, we rederive the equations of motion of MGN and verify that the MGN solution is only possible when the $g_{rr}$ component is a constant.

\section{Equations of motion and the dilaton black hole solution}\label{action}

We will be working with Einstein gravity minimally coupled to a scalar field described by the action (in units where $c=1$)
\begin{align}
 I&=\frac{1}{16\pi G}\int\dif^px\sqrt{|g|}\brac{R-\brac{\nabla\varphi}^2-\Vcal(\varphi)},
\end{align}
where $p$ is the number of spacetime dimensions. We will keep $p$ arbitrary in most of our equations here, though our explicit examples will be motivated by astrophysical contexts, in which case $p=4$.

Extremising the action gives the Einstein-dilaton equations
\begin{align}
 R_{\mu\nu}&=\nabla_\mu\varphi\nabla_\nu\varphi+\frac{1}{p-2}\Vcal g_{\mu\nu},\label{EOM_Einstein}\\
 \nabla^2\varphi&=\half\frac{\dif\Vcal}{\dif\varphi}.\label{EOM_dilaton}
\end{align}
We will consider the following dilaton potential:
\begin{align}
 \Vcal&=2V_0\expo{2b\varphi}.
\end{align}
Such a potential allows the Einstein-dilaton equation to be solved exactly. One solution of our present interest is the black hole\cite{Chan:1995fr}
\begin{subequations}\label{dbh_soln}
\begin{align}
 \dif s^2&=-f\dif t^2+h\dif r^2+r^2\dif\Sigma^2_{(p-2)},\\
        f&=\frac{r^{2a^2}}{h}=\frac{k(p-3)}{(p-3+a^2)(1-a^2)}r^{2a^2}-\mu r^{-(p-3-a^2)},\\
        \varphi&=-a\sqrt{p-2}\ln r+\varphi_0,\\
        2V_0&\expo{2b\varphi_0}=-\frac{(p-3)(p-2)ka^2}{(p-3+a^2)(1-a^2)},\quad b=\frac{1}{\sqrt{p-2}a}.\label{dbh_V}
\end{align}
\end{subequations}
Here, $\dif\Sigma^2_{(p-2)}=\tilde{\gamma}_{ij}\dif\sigma^i\dif\sigma^j$ denotes a maximally symmetric space with constant curvature $k=\pm 1$, $0$. The continuous parameter $\mu$ parametrises the mass of the black hole. In particular, by applying the appropriate counter-term subtraction on the boundary stress tensor of the spacetime \cite{Cai:1999xg}, the mass is found to be 
\begin{align}
 M=\frac{(p-2)\Sigma}{16\pi G}\mu,
\end{align}
where $\Sigma=\int\dif^{p-2}x\sqrt{\tilde{\gamma}}$. The scalar field is given up to an arbitrary integration constant $\varphi_0$ which we can set to zero without loss of generality. Another parameter characterising the spacetime is $a$, which determines the strength of the scalar field $\varphi$. The remaining quantities, $b$ and $V_0$ are fixed by $a$ via Eq.~\Eqref{dbh_V}. Indeed, for $a=0$, the solution reduces to the Schwarzschild black hole with $f=\frac{1}{h}=1-\frac{\mu}{r^{p-3}}$. 

This solution contains the spacetime which satisfies the \emph{flat curve condition} considered by MGN \cite{Matos:2000ki}. To recover it we simply set $\mu=0$, $p=4$, $a=\sqrt{l/2}$ and $k=1$, where the solution reduces to
\begin{subequations}\label{MGsolution}
\begin{align}
 \dif s^2&=-B_0r^{l}\dif t^2+h\dif r^2+r^2\brac{\dif\theta^2+\sin^2\theta\,\dif\phi^2}, \label{MatosMetric}\\
   h&=\frac{4-l^2}{4},\quad B_0=\frac{4}{4-l^2},\\
 \varphi&=-\sqrt{l}\ln r+\varphi_0. \label{MatosPhi}
\end{align}
\end{subequations}
This is precisely the solution studied in Ref.~\cite{Matos:2000ki} for $a^2=l/2$. Nandi et al. \cite{Nandi:2009hw} subsequenty considered the metric \Eqref{MatosMetric}, but with 
\begin{align}
 h=\frac{4-l^2}{4+C(4-l^2)r^{-(l+2)}}, \label{NandiMetric}
\end{align}
which is intended as a generalisation of the study by \cite{Matos:2000ki} for $C\neq0$. However, Eq.~\Eqref{NandiMetric} cannot be a solution to the Einstein equations if the scalar field is given by \Eqref{MatosPhi}. We demonstrate this explicitly in Appendix \ref{MGN}.

Finally, we write down its corresponding stress-energy tensor,
\begin{align}
 8\pi G T_{\mu\nu}&=\nabla_\mu\varphi\nabla_\nu\varphi-\half\brac{\nabla\varphi}^2g_{\mu\nu}-\half\Vcal g_{\mu\nu}.
\end{align}
Taking the stress-energy tensor to be the form of the perfect fluid, the energy density is $-{T^t}_t=\rho=-{T^i}_j$ and the pressures is ${T^r}_r=p$, where 
\begin{align}
 8\pi G\rho&=\half\sbrac{-a^2(p-2)fr^{-2a^2-2}+\frac{k(p-3)(p-2)a^2}{(p-3+a^2)(1-a^2)}r^{-2}},\\
 8\pi Gp&=\half\sbrac{a^2(p-2)fr^{-2a^2-2}+\frac{k(p-3)(p-2)a^2}{(p-3+a^2)(1-a^2)}r^{-2}}.
\end{align}

\section{Geodesic equations} \label{geodesics}

Having established the background spacetime, we now turn to the problem of particle motion in the solution. We parametrise the trajectory of the time-like or null particle by $x^\mu(\tau)$, where $\tau$ is an appropriate affine parametrisation. The trajectory will be a solution that extremises the Lagrangian $\Lcal=\half g_{\mu\nu}\dot{x}^\mu\dot{x}^\nu$, where over-dots denote derivatives with respect to $\tau$. We choose the affine parameter $\tau$ such that the invariant $g_{\nu\nu}\dot{x}^\mu\dot{x}^\nu=\epsilon$ has unit magnitude if it is non-zero. Therefore $\epsilon$ encodes the type of geodesic under consideration where 
\begin{align}
 \epsilon=\left\{\begin{array}{rl}
                  -1 & \mbox{for time-like particles}\\
                   0 & \mbox{for null/light-like particles}.
                 \end{array}\right.
\end{align}
For simplicity, let us also assume that the motion in $\dif\Sigma^2_{(p-2)}=\tilde{\gamma}_{ij}\dif\sigma^i\dif\sigma^j$ is suppressed in all but one direction, so that $\dot{\sigma}^i=(\dot{\phi},0,\ldots,0)$.

With these considerations, the Lagrangian has the explicit form
\begin{align}
 \Lcal=\half\brac{-f\dot{t}^2+h\dot{r}^2+r^2\dot{\phi}^2}.
\end{align}
Since $t$ and $\phi$ are cyclic variables in the Lagrangian, we have the first integrals
\begin{align}
 \dot{t}=\frac{E}{f},\quad\dot{\phi}=\frac{L}{r^2}, \label{dotphi}
\end{align}
where $E$ and $L$ are constants of motion which we may interpret as the particle's energy and angular momentum, respectively. Substituting these into $g_{\mu\nu}\dot{x}^\mu\dot{x}^\nu=\epsilon$ provides another first integral equation 
\begin{align}
 \dot{r}^2+\mathcal{U}=0, \label{FirstIntegral}
\end{align}
where we have defined the effective potential
\begin{align}
 \Ucal&=-r^{-2a^2}\sbrac{E^2-\brac{\frac{L^2}{r^2}-\epsilon}f}. \label{eqn_U}
\end{align}
The existence of an orbit requires $\dot{r}$ to be real, or equivalently, $\Ucal\leq0$. Therefore, for a given choice of constants, the condition $\Ucal\leq0$ determines the allowed range of $r$ for a particular orbit.

Another equation of motion for $r$ can be obtained by applying the Euler-Lagrange equation, $\frac{\dif}{\dif\tau}\frac{\partial\Lcal}{\partial\dot{r}}=\frac{\partial\Lcal}{\partial r}$, giving 
\begin{align}
\ddot{r}&=-\frac{h'}{2h}\dot{r}^2-\frac{f'E^2}{2f^2h}+\frac{L^2}{hr^3}. \label{ddotr}
\end{align}
where primes denote derivatives with respect to $r$. 

Our procedure of obtaining numerical solutions is as follows: For a choice of spacetime parameters $(\mu,a)$ and orbital constants $(E,L)$, Eqs.~\Eqref{ddotr} and \Eqref{FirstIntegral} are integrated using the fourth-order Runge-Kutta method, with the aid of \Eqref{eqn_U} to serve as a consistency check of the results. The initial conditions we typically adopt is $\dot{r}=0$, with initial $r$ calculated by substituting $\dot{r}=0$ into \Eqref{eqn_U}. For concreteness, in all the following examples we consider spherically symmetric dilaton black hole in four dimensions where $k=1$ and $p=4$.

\begin{figure}[t]
 \begin{center}
  \begin{subfigure}[b]{\textwidth}
   \centering
   \includegraphics[scale=0.8]{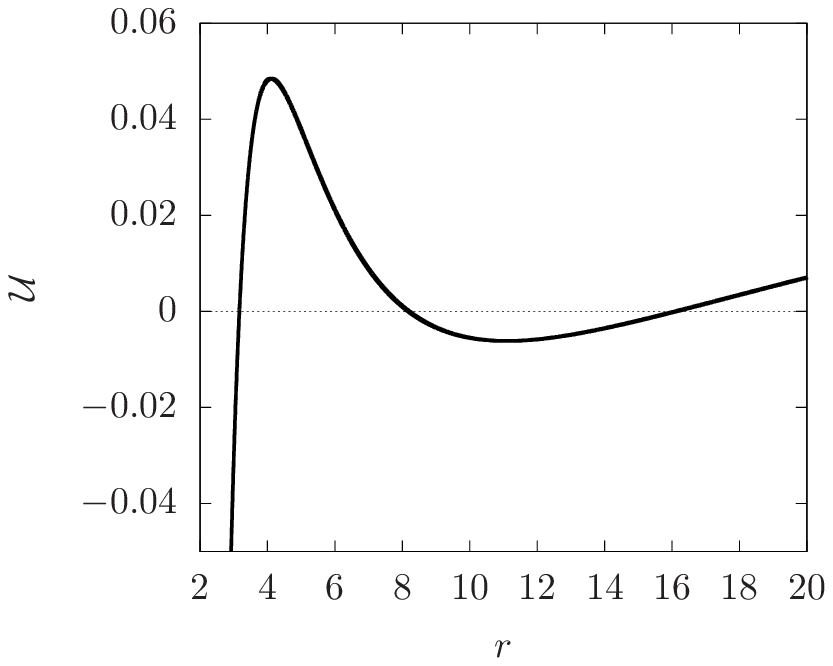}
   \includegraphics[scale=0.8]{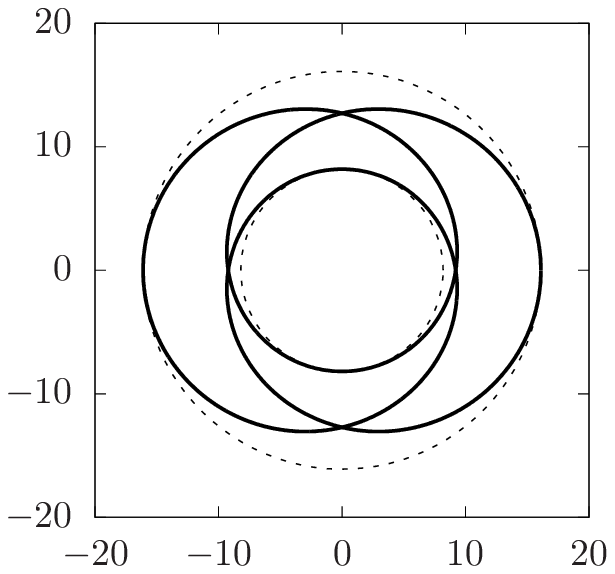}
   \caption{$a=0$.}
   \label{fig_PrelimDemo_a}
  \end{subfigure}
  
  \begin{subfigure}[b]{\textwidth}
   \centering
   \includegraphics[scale=0.8]{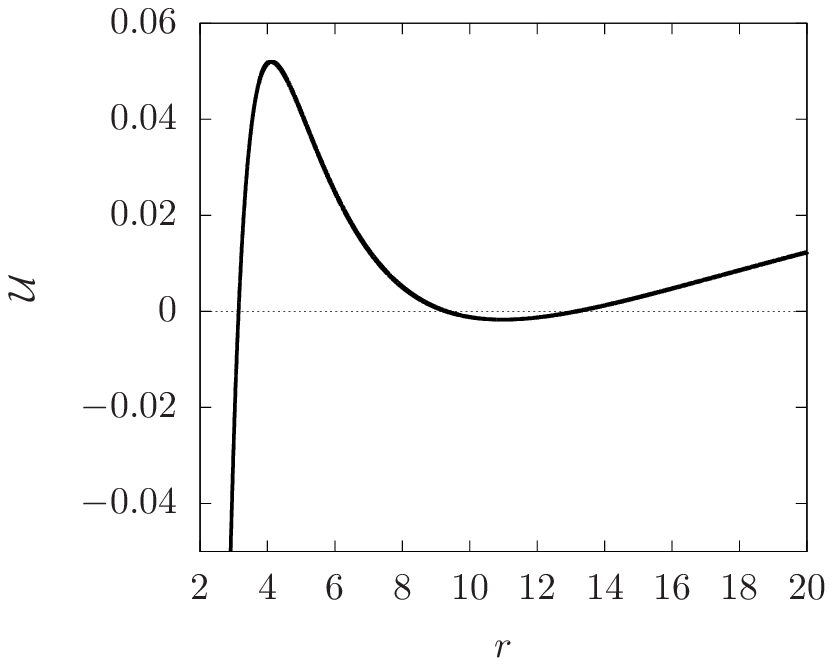}
   \includegraphics[scale=0.8]{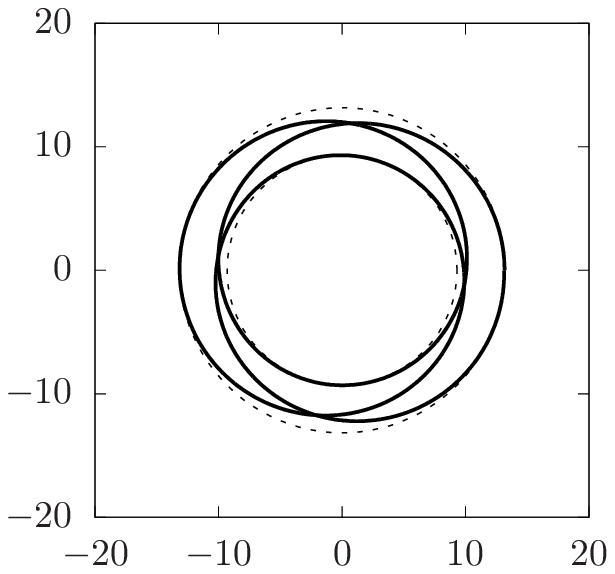}
   \caption{$a=0.03$.}
   \label{fig_PrelimDemo_b}
  \end{subfigure}
  
  \caption{(Colour online) Plots of effective potential against $r$ along with their corresponding orbits with energy $E=0.962903$ and $L=3.9$, in the four-dimensional spherically symmetric black hole spacetime with $p=4$, $k=1$, and $\mu=2$. The numerical solution shown here are integrated up to three periods of $\phi$.}
  \label{fig_PrelimDemo}

 \end{center}

\end{figure}

As a warm-up example of this procedure, consider the following representative orbits for the Schwarzschild solution $a=0$, $\mu=2$ in Fig.~\ref{fig_PrelimDemo_a} and the case $a=0.03$, $\mu=2$ in Fig.~\ref{fig_PrelimDemo_b}. for the choice\footnote{This particular $E$ and $L$ for $a=0$ corresponds to a closed Schwarzschild orbit labelled $(2,0,1)$ in \cite{Levin:2008mq}.} of $E=0.962903$ and $L=3.9$. The left panels for each case are plots of $\Ucal$ against $r$. As mentioned above, the condition $\Ucal\leq0$ determines the allowed range of $r$ for a trajectory. For the case $a=0$ as shown in Fig.~\ref{fig_PrelimDemo_a}, this is approximately $8.190\leq r\leq 16.109$, and for \ref{fig_PrelimDemo_b}, approximately $9.311\leq r\leq13.164$ for the case $a=0.03$. 

\section{Circular orbits and rotation curves of time-like orbits} \label{curves}

In this section we shall consider circular orbits of time-like and null particles. These orbits are characterised by the property $\dot{r}=0$. If $r_0$ is the radius of the orbit, then it is required that $\Ucal(r_0)=\Ucal'(r_0)=0$. For these orbits to be stable we also require $\Ucal''(r_0)>0$. Explicitly, the derivatives of the effective potential are 
\begin{align}
 \Ucal'&=-\frac{2a^2}{r}\Ucal-r^{-2a^2}\brac{\frac{2f-rf'}{r^3}L^2+\epsilon f'},\label{eqn_Up}\\
 \Ucal''&=\frac{2a^2(1-2a^2)}{r^2}\Ucal-\frac{4a^2}{r}\Ucal'-r^{-2a^2}\brac{\frac{4rf'-r^2f''-6f}{r^4}L^2+\epsilon f''}.\label{eqn_Upp}
\end{align}
For time-like particles, we have $\epsilon=-1$. Thus $\Ucal=\Ucal'=0$ leads to
\begin{align}
 E^2&=\frac{2f^2}{2f-rf'},\quad L^2=\frac{r^3f'}{2f-rf'},\label{circ_EL}
\end{align}
giving the required energy and angular momentum for a circular orbit of radius $r$. 

The stability of a circular orbit depends on the sign of $\Ucal''$. Substituting $\Ucal'=\Ucal=0$ and \Eqref{circ_EL} into \Eqref{eqn_Upp}, we have  
\begin{align}
 \Ucal''&=-r^{-2a^2}\brac{\frac{(4rf'-r^2f''-6f)f'}{r(2f-rf')}-f''}. \label{circ_Upp}
\end{align}
Typically, this is positive down to some lower bound of $r$, below which we only have unstable circular orbits with $\Ucal''<0$. This lower bound $r_{\mathrm{ISCO}}$ represents the critical radius which is called the \emph{innermost stable circular orbit} (ISCO), and we can determine its value by equating Eq.~\Eqref{circ_Upp} to zero and solving for $r$. 

\begin{figure}
 \begin{center}
  \includegraphics{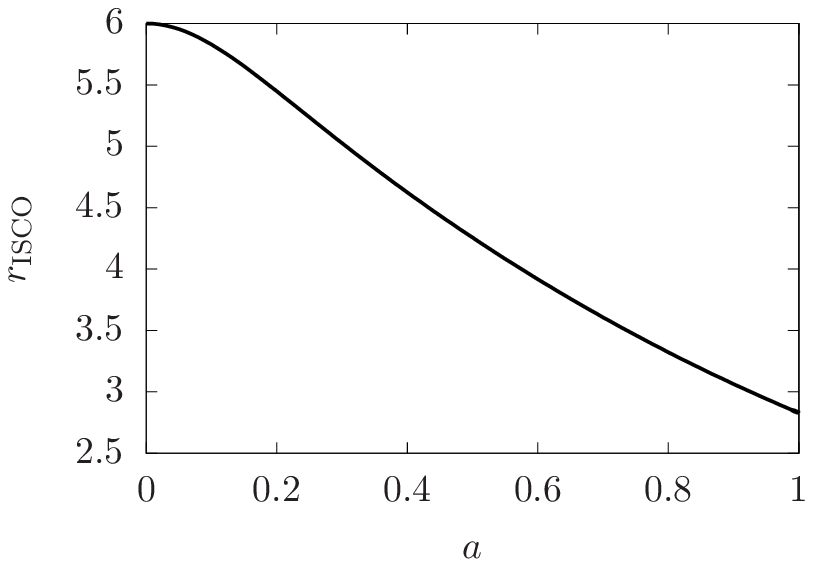}
  \caption{Plot of $r_{\mathrm{ISCO}}$ against $a$.}
  \label{fig_ISCO}
 \end{center}

\end{figure}

For the Schwarzschild case ($a=0$), this gives the well-known exact value of $r=3\mu$. For other values of dilaton parameter $a$ we obtain $r_{\mathrm{ISCO}}$ numerically. As shown in Fig.~\ref{fig_ISCO}, as $a$ is increased from zero, the value of $r_{\mathrm{ISCO}}$ decreases. In other words, the presence of the scalar field allows smaller stable circular orbits, at least naively in terms of the coordinate radius $r$. 

Turning now to the tangential velocity for these circular orbits, various definitions that was used in the literature. Here, let us use the definition where the velocity is extracted from the spatial part of $g_{\mu\nu}\dot{x}^\mu\dot{x}^\nu=\epsilon$. The expression for tangential velocity is\footnote{There is another definition proposed by \cite{Dey:2014gka} where $v_{\mathrm{tan}}=\sqrt{\half rf'}$.} \cite{Matos:2000jr} 
\begin{align}
 v_{\mathrm{tan}}=\frac{r}{\sqrt{f}}\frac{\dif\phi}{\dif t}=\sqrt{\frac{rf'}{2f}},
\end{align}
It is this definition that was used by \cite{Matos:2000ki} to determine the flat curve condition which lead to the solution \Eqref{MGsolution}. In that case we have a constant $v_{\mathrm{tan}}=\sqrt{2}a$, which was equivalent to $\sqrt{l}$ in the notation of Ref.~\cite{Matos:2000ki}. As long as $\mu=0$, this result is actually independent of of the dimensionality of the spacetime.

\begin{figure}
 \begin{center}
  \begin{subfigure}[b]{0.49\textwidth}
  \includegraphics[scale=0.85]{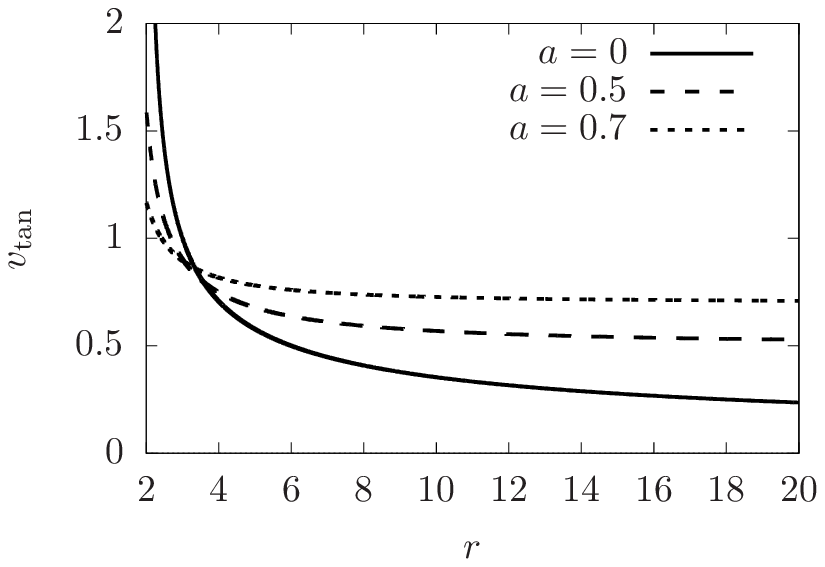}
  \caption{$\mu=2$, various $a$.}
  \label{fig_RotCurves_a}
  \end{subfigure}
  \begin{subfigure}[b]{0.49\textwidth}
  \includegraphics[scale=0.85]{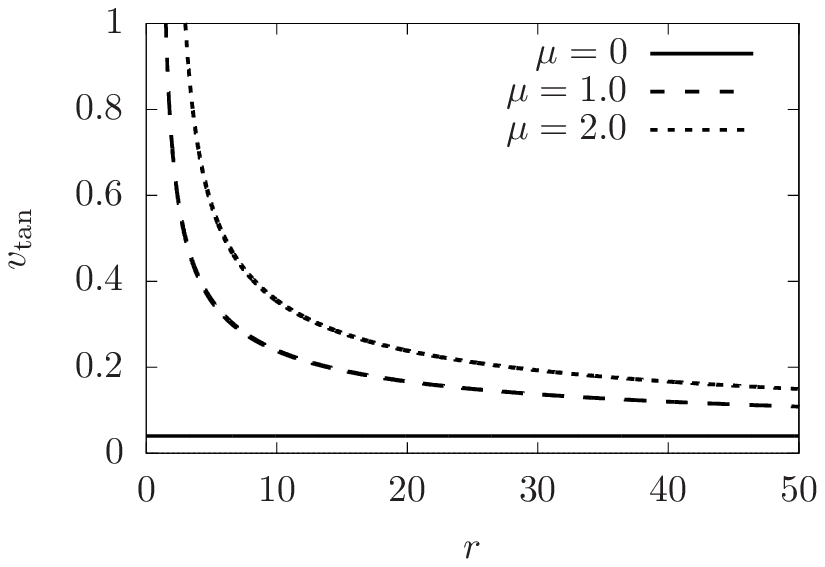}
  \caption{$a=0.04$, various $\mu$.}
  \label{fig_RotCurves_b}
  \end{subfigure}
  \caption{Plots of $v_{\mathrm{tan}}$ vs $r$, for (a) fixed $\mu$ and different $a$, and (b) fixed  }
  
 \end{center}

\end{figure}

The tangential velocities for non-zero $\mu$ and various $a$ are shown in Fig.~\ref{fig_RotCurves_a}, where the result is perhaps unsurprising. The presence of $\mu$ causes $v_{\mathrm{tan}}$ to be higher near the origin, where the gravitating mass is located. Further away, the velocity approaches $\sqrt{2}a$, as expected from the flat curve condition. Nevertheless, for small $r$, it is interesting to note that larger $a$ corresponds to lower speeds compared to curves of smaller $a$, indicating that the presence of the scalar field diminishes the gravitational attraction of the central mass. Also, at first glance these distances, this suggests the possibility of observable consequences for stars near the galactic core. But the values of $a$ we are using in Fig.~\ref{fig_RotCurves_a} are quite large. Based on the observed tangential velocities of galaxies, it is required that $v_{\mathrm{tan}}\sim\sqrt{2}a\sim 10^{-3}$ (in units of $c$).

\section{Photon spheres and gravitational lensing} \label{lensing}

\begin{figure}
 \begin{center}
  \begin{subfigure}[b]{0.49\textwidth}
   \includegraphics[scale=0.79]{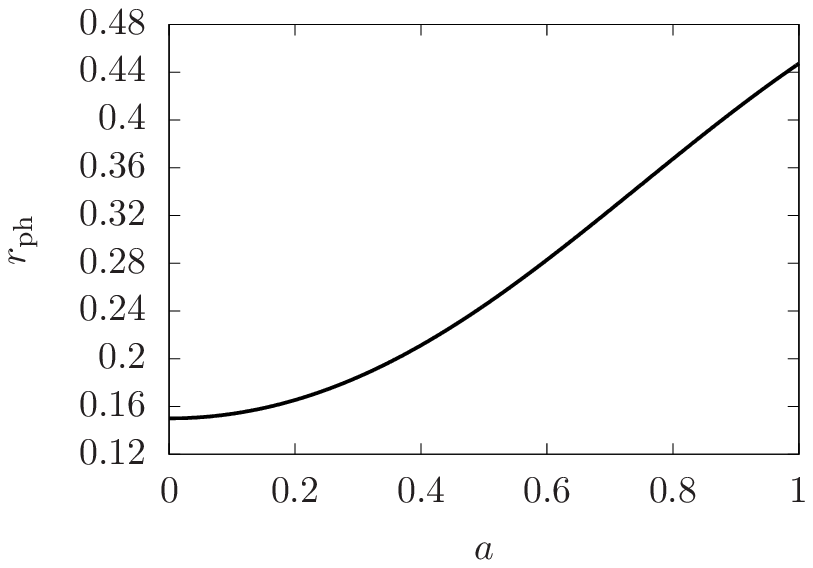}
   \caption{$\mu=0.1$.}
   \label{fig_PhotonSphere_a}
  \end{subfigure}
  \begin{subfigure}[b]{0.49\textwidth}
   \includegraphics[scale=0.79]{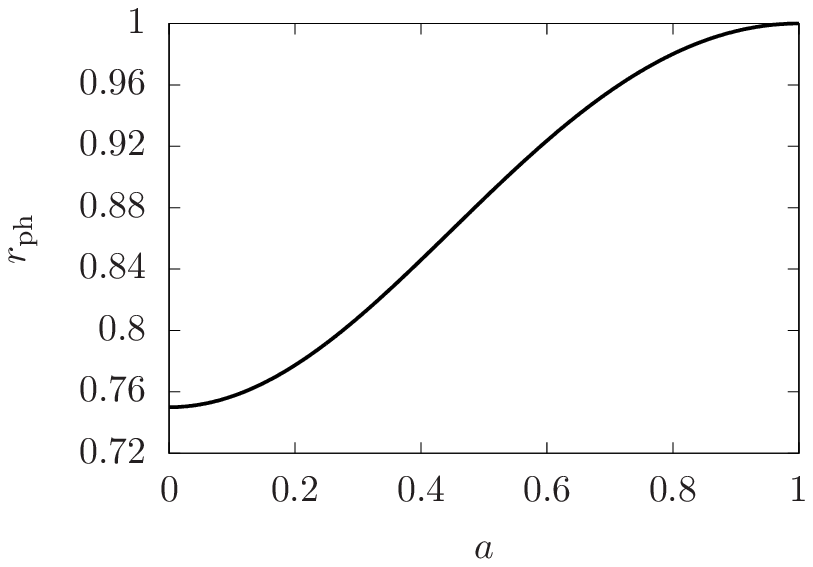}
   \caption{$\mu=0.5$.}
   \label{fig_PhotonSphere_b}
  \end{subfigure}
  \begin{subfigure}[b]{0.49\textwidth}
   \includegraphics[scale=0.79]{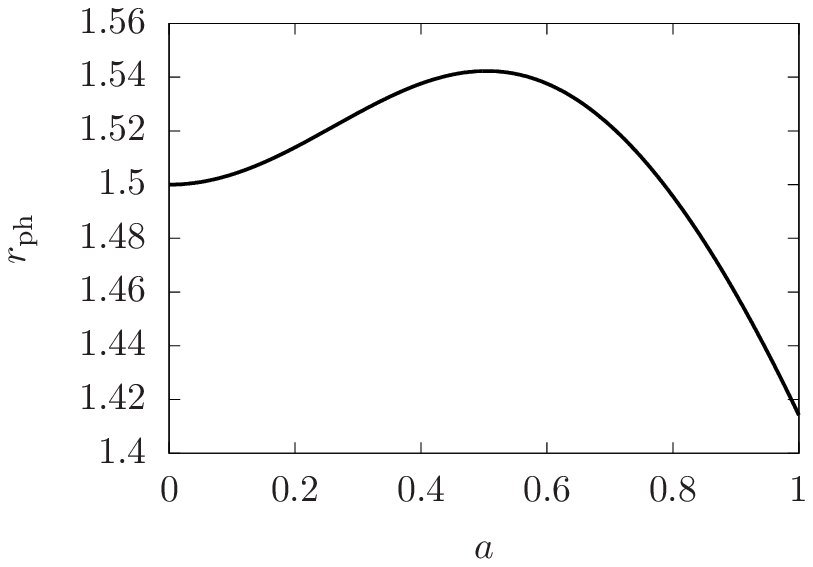}
   \caption{$\mu=1.0$.}
   \label{fig_PhotonSphere_c}
  \end{subfigure}
  \begin{subfigure}[b]{0.49\textwidth}
   \includegraphics[scale=0.79]{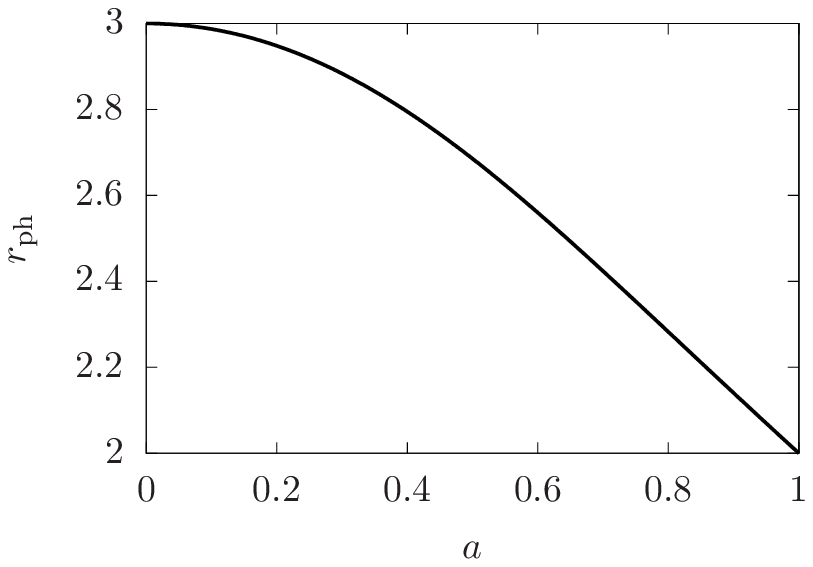}
   \caption{$\mu=2.0$.}
   \label{fig_PhotonSphere_d}
  \end{subfigure}
  \caption{Plots of photon sphere radii against dilaton parameter $a$ for various black hole masses.}
  \label{fig_PhotonSphere}
 \end{center}

\end{figure}

We now consider orbits for photon trajectories with $\epsilon=0$. Looking for circular orbits, the condition $\Ucal=0$ gives $\frac{L^2}{E^2}=\frac{r^2}{f}$, which is the standard result of the impact parameter at the distance of closest approach. On the other hand, the condition $\Ucal'=0$ gives 
\begin{align}
 2f&=rf'\quad\rightarrow\quad r_{\mathrm{ph}}=\sbrac{\half(3-a^2)(1+a^2)\mu}^{\frac{1}{1+a^2}}, \label{r_ph}
\end{align}
where $r_{\mathrm{ph}}$ is the radius of the photon sphere. Clearly, the case $a=0$ reduces to the standard Schwarzschild result of  $r_{\mathrm{ph}}=\frac{3}{2}\mu$. The radii for other non-zero values of $a$ are shown in Fig.~\ref{fig_PhotonSphere}. We see that starting small $\mu$, for instance $\mu=0.1$ in Fig.~\ref{fig_PhotonSphere_a}, a non-zero $a$ increases the size of the photon sphere. If $\mu$ is relatively larger, say, $\mu=1$ in Fig.~\ref{fig_PhotonSphere_c}, the radii develops a maximum at some intermediate $a$. The location of the maximum decreases with increasing $\mu$. At larger $\mu$, the the largest photon sphere is for $a=0$, in which case all photon spheres of non-zero $a$ are smaller than the Schwarzschild case. Evaluating the second derivative of $\Ucal$ at the photon sphere tells us that $\Ucal''(r_{\mathrm{ph}})<0$ in all cases. Therefore all photon spheres are unstable. 

Next let us consider gravitational lensing in this spacetime. Since the spacetime we are considering is not flat, we are not able to consider the deflection of light which comes from $r\rightarrow\infty$. Nevertheless, we can consider photon trajectories that begin and end at finite $r$. Since the spacetime is not flat at finite distances from the black hole, we should be careful not to take the changes in the coordinate value $\phi$ as the bending angle. 

\begin{figure}
 \begin{center}
  \includegraphics{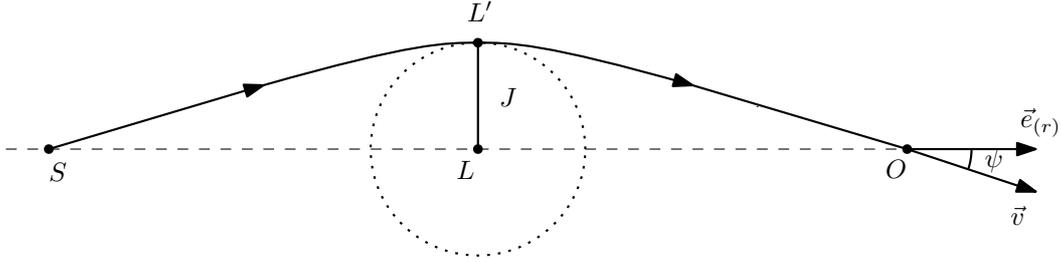}
  \caption{Schematic diagram depicting gravitational lensing in the spacetime of a dilaton black hole. The trajectory begins at the source $S$ and is deflected by the dilaton black hole lens $L$ at impact parameter $J=r_0/\sqrt{f(r_0)}$. Then the photon proceeds towards the observer at the intersection of the trajectory and the optic axis.}
  \label{fig_LensDiagram}
 \end{center}

\end{figure}

However, we can adopt the Rindler-Ishak method \cite{Rindler:2007zz} to establish a coordinate invariant quantity which describes the deflection of the photon trajectory. For simplicity, we shall consider the case where the source $S$, lens $L$ and observer $O$ are co-aligned on the optic axis, as shown in Fig.~\ref{fig_LensDiagram}.. We shall call the point of closest approach $L'$, which has coordinate value $r_0$. 

\begin{figure}
 \begin{center}
 \begin{subfigure}[b]{0.49\textwidth}
  \includegraphics[scale=0.86]{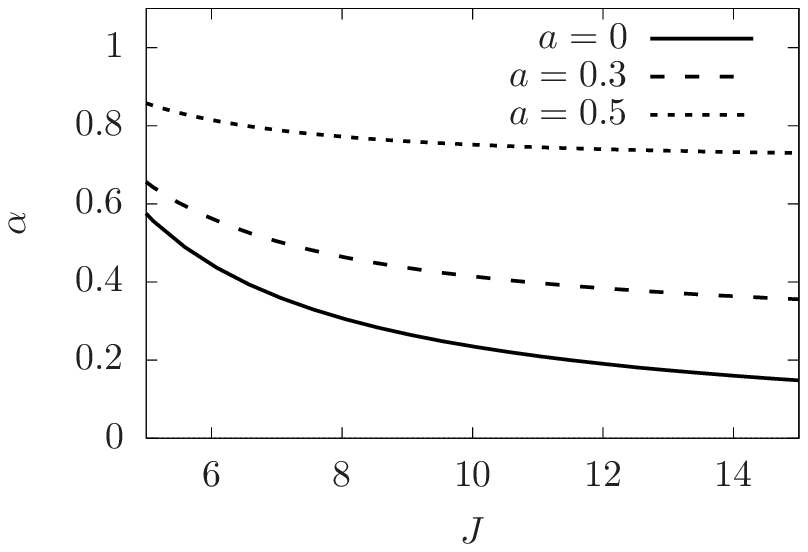}
  \caption{$\mu=1$.}
  \label{fig_BendingAngle_a}
 \end{subfigure}
 \begin{subfigure}[b]{0.49\textwidth}
  \includegraphics[scale=0.86]{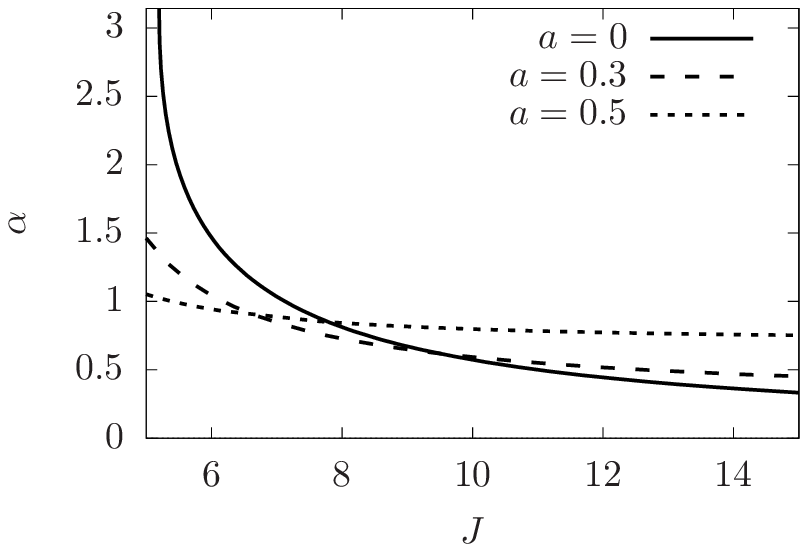}
  \caption{$\mu=2$.}
  \label{fig_BendingAngle_b}
 \end{subfigure}
  \caption{Plots of bending angle $\alpha$ vs impact parameter $J$ for $\mu=1,2$, and various $a$.}
  \label{fig_BendingAngle}
 \end{center}

\end{figure}

In the Rindler-Ishak method, we establish a spatial ortho-normal frame at the location of the observer $O$. In the two-dimensional plane containing the null geodesic, it is given by
\begin{align}
 \vec{e}_{(r)}=\frac{1}{\sqrt{h}}\partial_r,\quad \vec{e}_{(\phi)}=\frac{1}{r}\partial_\phi.
\end{align}
We take the spatial part of the photon's four-velocity $(\dot{t},\vec{v})$, where $\vec{v}=g_{ij}\dot{x}^i\dot{x}^j$, where $i,j\neq t$, as $\vec{v}=\dot{r}\partial_r+\dot{\phi}\partial_\phi$. An invariant angle $\psi$ is given by
\begin{align}
 \cos\psi=\vec{e}_{(r)}\cdot\frac{\vec{v}}{|\vec{v}|},
\end{align}
and is depicted as the angle between the direction of the photon and the optic axis at $O$ in Fig.~\ref{fig_LensDiagram}. Using the geodesic equations, we can show that
\begin{align}
 \sin\psi=\frac{r_0}{r}\frac{\sqrt{f(r)}}{\sqrt{f(r_0)}},
\end{align}
where $r=r_0$ is the coordinate distance of closest approach. The total deflection of the trajectory beginning with $S$ and ending at $O$ is twice of this angle,
\begin{align}
 \hat{\alpha}=2\psi=2\sin^{-1}\brac{\frac{r_0}{r}\frac{\sqrt{f(r)}}{\sqrt{f(r_0)}}}.\label{alphahat}
\end{align}
We can parametrise each photon trajectory by $r_0$ with its corresponding impact parameter $J=r_0/\sqrt{f(r_0)}$. By the spherical symmetry of the problem, it suffices to calculate $\hat{\alpha}$ by integrating Eqs.~\Eqref{dotphi} and \Eqref{ddotr} starting from point $L'$ with initial conditions $r=r_0$, $\dot{r}=0$, and $\phi=0$. The point $O$ where the trajectory intersects the optic axis is $\phi=\frac{\pi}{2}$. The value of $r$ at this point is then substituted into \Eqref{alphahat}. The results of the integration are shown in Fig.~\ref{fig_BendingAngle}.

Between Fig.~\ref{fig_BendingAngle_a} and \ref{fig_BendingAngle_b}, we see that increasing $a$ raises the value of the bending angle uniformly across the range of impact parameter, while $\mu$, reflecting the central black-hole mass, causes a large bending angle only for impact parameter close to the black hole, as we expect intuitively. However, note that for larger $\mu$ (say, $\mu=2$ as shown in Fig.~\ref{fig_BendingAngle_b}), the bending angles for $J$ close to the black hole are smaller for larger $a$, showing us that the presence of the scalar field diminishes the gravitational effect of the central mass, just as in the time-like geodesic case.

\section{Conclusion} \label{conclusion}

In this paper we have derived and studied the geodesic equations of the dilaton black hole with an exponential potential. Since the zero mass limit of this solution corresponds to the MGN spacetime, we may treat particle motion around the dilaton black hole spacetime as a toy model generalising the MGN solution to include a central gravitating mass. 

From the rotation curves of time-like particles and gravitational lensing of photon trajectories, we find that the presence of the dilaton scalar field roughly increases the overall gravitational effects for all distances. On the other hand, at distances close to the origin where we expect gravitational attraction to be strong, the presence of the dilaton diminishes is effect, as can be seen from the lower tangential velocities and smaller bending angle at small distances compared to the Schwarzschild case. These diminishing effects at small distances clearly not consistent with observations, since we do not expect deviations from standard gravity at sub-galactic distances or large accelerations. However, the case where the mass zero or sufficiently small approximates the metric obtained by MGN which was designed to satisfy the flat curve condition. 

Looking at these ranges, the results of Sec.~\ref{lensing} shows that the bending angle is larger compared to the Schwarzschild case of the same mass, and thus the field $\varphi$ plays the role of dark matter which causes an additional bending of light on top of those from baryonic mass. However, we shall stop short of making these claims in the context of actual observations, since, as mentioned above, our spacetime is a highly idealised model which does not capture all the subtleties of galactic and cosmological dynamics.

\appendix 

\section{Equations of motion for the MGN solution} \label{MGN}

In this Appendix we shall rederive the MGN metric and show that the solution with non-zero integration constant considered by \cite{Nandi:2009hw} cannot be an exact solution to the Einstein-dilaton equations. We begin by taking the metric ansatz to be
\begin{align}
 \dif s^2&=-f\dif t^2+h\dif r^2+r^2\dif\Sigma^2_{(p-2)},
\end{align}
where $f$ and $h$ are functions that depend only on the coordinate $r$. Assuming the same for the scalar field $\varphi$,  the Einstein equations are
\begin{subequations}\label{EE}
\begin{align}
 -\frac{1}{2\sqrt{fh}}\brac{\frac{f'}{\sqrt{fh}}}'-\frac{(p-2)f'}{2rfh}&=\frac{1}{p-2}\Vcal,  \label{EE1}\\
 -\frac{1}{2\sqrt{fh}}\brac{\frac{f'}{\sqrt{fh}}}'+\frac{(p-2)h'}{2rh^2}&=\frac{\varphi'^2}{h}+\frac{1}{p-2}\Vcal,  \label{EE2}\\
 \frac{p-3}{r^2h}\brac{kh-1}+\frac{h'}{2rh^2}-\frac{f'}{2rfh}&=\frac{1}{p-2}\Vcal.  \label{EE3}
\end{align}
\end{subequations}
In this section, primes denote derivatives with respect to $r$. For completeness, we write down the dilaton equation,
\begin{align}
 \frac{1}{\sqrt{fh}r^{p-2}}\brac{\sqrt{fh}r^{p-2}h^{-1}\varphi'}'&=\half\frac{\dif\Vcal}{\dif\varphi}. \label{d}
\end{align}
Eqs.~\Eqref{EE} and \Eqref{d} may serve as a starting point to derive the solution \Eqref{dbh_soln}. 

However, we are presently interested in verifying the solution by MGN \cite{Matos:2000ki}. To do this we first take the difference between \Eqref{EE1} and \Eqref{EE2} to obtain
\begin{align}
 \frac{h'}{2rh^2}+\frac{f'}{2rfh}=\frac{1}{p-2}\frac{\varphi'^2}{h}.\label{fh_eqn}
\end{align}
Using this to remove $f'$ from \Eqref{EE3} gives
\begin{align}
 \frac{1}{r^2}\brac{(p-3)(1-kh)-r\frac{h'}{h}}&=-\brac{\frac{1}{p-2}\varphi'^2+\frac{1}{p-2}\Vcal h}. \label{preMatos1}
\end{align}
On the other hand, if we used \Eqref{fh_eqn} to remove $h'$ from \Eqref{EE3} instead, we would have 
\begin{align}
 \frac{1}{r^2}\brac{(p-3)(kh-1)-r\frac{f'}{f}}&=-\brac{\frac{1}{p-2}\varphi'^2-\frac{1}{p-2}\Vcal h}. \label{preMatos2}
\end{align}
Finally, we note that since
\begin{align}
 \frac{\varphi'^2}{h}&=\frac{1}{p-2}\frac{\varphi'^2}{h}+\frac{p-3}{p-2}\frac{\varphi'^2}{h}=\frac{h'}{2rh^2}+\frac{f'}{2rfh}+\frac{p-3}{p-2}\frac{\varphi'^2}{h},
\end{align}
where Eq.~\Eqref{fh_eqn} has been used to take the second equality to the third. Substituting this into Eq.~\Eqref{EE2} gives 
\begin{align}
 \frac{1}{4r^2}\sbrac{-2r^2\frac{f''}{f}+r^2\frac{f'^2}{f^2}-2r\frac{f'}{f}+\brac{2(p-3)+r\frac{f'}{f}}r\frac{h'}{h}}&=\frac{p-3}{p-2}\varphi'^2+\frac{1}{p-2}\Vcal h.\label{preMatos3}
\end{align}
If we substitute $k=1$, $p=4$, $\Vcal=2V$, and $f=r^l$, Eqs.~\Eqref{preMatos1}, \Eqref{preMatos2}, and \Eqref{preMatos3} becomes\footnote{Obtaining Eq.~\Eqref{MatosEqns} is more direct if we had started with the Einstein equation in the `Einstein tensor form', $R_{\mu\nu}-\half Rg_{\mu\nu}=\nabla_\mu\varphi\nabla_\nu\varphi-\half\brac{\nabla\varphi}^2-\half\Vcal g_{\mu\nu}$. Then Eqs.~\Eqref{Matos3}--\Eqref{Matos2} are simply the $tt$, $rr$, and $ij$-components.} 
\begin{subequations}\label{MatosEqns}
\begin{align}
 \frac{1}{r^2}\brac{1-h-r\frac{h'}{h}}&=-\brac{\half\varphi'^2+\half\Vcal h},\label{Matos3}\\
 \frac{1}{r^2}\brac{h-(l+1)}&=-\brac{\half\varphi'^2-\half\Vcal h},\label{Matos1}\\
 \frac{1}{4r^2}\brac{l^2+(2+l)r\frac{h'}{h}} &=-\brac{\half\varphi'^2+\half\Vcal h},\label{Matos2}
\end{align}
\end{subequations}
thus recovering Eqs.~(11)--(13) of \cite{Matos:2000ki} precisely.\footnote{Note that our scalar field $\varphi$ is defined such that the Newton's constant is absorbed. Namely the scalar field $\Phi$ in \cite{Matos:2000ki} is related to $\varphi$ via $\Phi=\frac{1}{\sqrt{\kappa_0}}\varphi$, where $\kappa_0=8\pi G$. Furthermore the potential $V$ in \cite{Matos:2000ki} is related to $\Vcal$ via $V=\frac{2}{\kappa_0}\Vcal$.} In Ref.~\cite{Matos:2000ki}, to obtain a spacetime that supports orbits consistent with the galactic rotation curves, the potential 
\begin{align}
 \half\Vcal=-\frac{l}{2-l}\expo{-2\varphi/\sqrt{l}}
\end{align}
was considered, which gives the solution  
\begin{align}
 h&=\frac{4-l^2}{4},\label{Matos_h}
\end{align}
The scalar field solution is
\begin{align}
 \varphi=\sqrt{l}\ln r+\varphi_0 \label{Matos_phi}
\end{align}
Subsequently, Ref.~\cite{Nandi:2009hw} considered the more general function
\begin{align}
 h&=\frac{4-l^2}{4+C(4-l^2)r^{-(l+2)}},\label{Nandi_h}
\end{align}
along with \Eqref{Matos_phi} as the scalar field. Ref.~\cite{Nandi:2009hw} investigated the physical properties of the galactic halo for a (small, but) non-zero $C$. However, we will now see that \Eqref{Nandi_h}, together with \Eqref{Matos_phi} cannot be an exact solution to \Eqref{MatosEqns} unless $C=0$, for which it reduces to the case \Eqref{Matos_h} originally considered in \cite{Matos:2000ki}. This is easily seen by direct substitution. For instance, substituting \Eqref{Nandi_h} into the left hand side of Eq.~\Eqref{Matos3} gives 
\begin{align}
 \frac{1}{r^2}\brac{1-h-r\frac{h'}{h}}&=\frac{1}{r^2}\brac{\frac{l^2-(l+1)C(4-l^2)r^{-(l+2)}}{4+C(4-l^2)r^{-(l+2)}}},
\end{align}
but substituting \Eqref{Matos_phi} into the right hand side of the same equation gives 
\begin{align}
 -\brac{\half\varphi'^2+\half\Vcal h}=\frac{1}{r^2}\brac{\frac{l^2-\half lC(4-l^2)r^{-(l+2)}}{4+C(4-l^2)r^{-(l+2)}}}.
\end{align}
Clearly, the two sides are not equal unless $C=0$. Similar checks can be done with Eqs.~\Eqref{Matos1} and \Eqref{Matos2}.

\bibliographystyle{dbh}

\bibliography{dbh}

\providecommand{\href}[2]{#2}\begingroup\raggedright\begin{thebibliography}{10}

\bibitem{Rubin:1980zd}
V.~C. Rubin, N.~Thonnard, and W.~K. Ford, Jr., {\it {`Rotational properties of
  21 SC galaxies with a large range of luminosities and radii, from NGC 4605 /R
  = 4kpc/ to UGC 2885 /R = 122 kpc/'}},  Astrophys.~J. {\bf 238} (1980) 471.

\bibitem{Sofue:2000jx}
Y.~Sofue and V.~Rubin, {\it {`Rotation curves of spiral galaxies'}},
  Ann.~Rev.~Astron.~Astrophys. {\bf 39} (2001) 137,
  [\href{http://arxiv.org/abs/astro-ph/0010594}{{\tt astro-ph/0010594}}].

\bibitem{Persic:1995ru}
M.~Persic, P.~Salucci, and F.~Stel, {\it {`The Universal rotation curve of
  spiral galaxies: 1. The Dark matter connection'}},
  Mon.~Not.~Roy.~Astron.~Soc. {\bf 281} (1996) 27,
  [\href{http://arxiv.org/abs/astro-ph/9506004}{{\tt astro-ph/9506004}}].

\bibitem{CabralRosetti:2001bh}
L.~G. Cabral-Rosetti, T.~Matos, D.~Nunez, and R.~A. Sussman, {\it
  {`Hydrodynamics of galactic dark matter'}},  Class.~Quant.~Grav. {\bf 19}
  (2002) 3603, [\href{http://arxiv.org/abs/gr-qc/0112044}{{\tt
  gr-qc/0112044}}].

\bibitem{Matos:2001pz}
T.~Matos and F.~S. Guzman, {\it {`On the space time of a galaxy'}},
  Class.~Quant.~Grav. {\bf 18} (2001) 5055,
  [\href{http://arxiv.org/abs/gr-qc/0108027}{{\tt gr-qc/0108027}}].

\bibitem{Matos:2000jr}
T.~Matos, D.~Nunez, F.~S. Guzman, and E.~Ramirez, {\it {`Geometric conditions
  on the type of matter determining the flat behavior of the rotational curves
  in galaxies'}},  Gen.~Rel.~Grav. {\bf 34} (2002) 283,
  [\href{http://arxiv.org/abs/astro-ph/0005528}{{\tt astro-ph/0005528}}].

\bibitem{Rahaman:2010xs}
F.~Rahaman, K.~K. Nandi, A.~Bhadra, M.~Kalam, and K.~Chakraborty, {\it
  {`Perfect Fluid Dark Matter'}},  Phys.~Lett.~B {\bf 694} (2011) 10,
  [\href{http://arxiv.org/abs/1009.3572}{{\tt arXiv:1009.3572}}].

\bibitem{Diez-Tejedor:2013sza}
A.~Diez-Tejedor and A.~X. Gonzalez-Morales, {\it {`No-go theorem for static
  scalar field dark matter halos with no Noether charges'}},  Phys.~Rev.~D {\bf
  88} (2013) 067302, [\href{http://arxiv.org/abs/1306.4400}{{\tt
  arXiv:1306.4400}}].

\bibitem{Dey:2014gka}
D.~Dey, K.~Bhattacharya, and T.~Sarkar, {\it {`Galactic space-times in modified
  theories of gravity'}},  Gen.~Rel.~Grav. {\bf 47} (2015) 103,
  [\href{http://arxiv.org/abs/1407.0319}{{\tt arXiv:1407.0319}}].

\bibitem{Matos:2000ki}
T.~Matos, F.~S. Guzman, and D.~Nunez, {\it {`Spherical scalar field halo in
  galaxies'}},  Phys.~Rev.~D {\bf 62} (2000) 061301,
  [\href{http://arxiv.org/abs/astro-ph/0003398}{{\tt astro-ph/0003398}}].

\bibitem{Nandi:2009hw}
K.~K. Nandi, I.~Valitov, and N.~G. Migranov, {\it {`Remarks On The Spherical
  Scalar Field Halo In Galaxies'}},  Phys.~Rev.~D {\bf 80} (2009) 047301,
  [\href{http://arxiv.org/abs/0907.4016}{{\tt arXiv:0907.4016}}]. [Erratum:
  Phys. Rev.D83,029902(2011)].

\bibitem{Chan:1995fr}
K.~C.~K. Chan, J.~H. Horne, and R.~B. Mann, {\it {`Charged dilaton black holes
  with unusual asymptotics'}},  Nucl.~Phys.~B {\bf 447} (1995) 441,
  [\href{http://arxiv.org/abs/gr-qc/9502042}{{\tt gr-qc/9502042}}].

\bibitem{Cai:1997ii}
R.-G. Cai, J.-Y. Ji, and K.-S. Soh, {\it {`Topological dilaton black holes'}},
  Phys.~Rev.~D {\bf 57} (1998) 6547,
  [\href{http://arxiv.org/abs/gr-qc/9708063}{{\tt gr-qc/9708063}}].

\bibitem{Charmousis:2001nq}
C.~Charmousis, {\it {`Dilaton space-times with a Liouville potential'}},
  Class.~Quant.~Grav. {\bf 19} (2002) 83,
  [\href{http://arxiv.org/abs/hep-th/0107126}{{\tt hep-th/0107126}}].

\bibitem{Dehghani:2004jk}
M.~H. Dehghani, {\it {`Magnetic strings in dilaton gravity'}},  Phys.~Rev.~D
  {\bf 71} (2005) 064010, [\href{http://arxiv.org/abs/hep-th/0411274}{{\tt
  hep-th/0411274}}].

\bibitem{Agop:2005np}
M.~Agop, E.~Radu, and R.~Slagter, {\it {`On Ernst black holes with a dilaton
  potential'}},  Mod.~Phys.~Lett.~A {\bf 20} (2005) 1077.

\bibitem{Abdolrahimi:2011zg}
S.~Abdolrahimi and A.~A. Shoom, {\it {`Geometric properties of static
  Einstein-Maxwell dilaton horizons with a Liouville potential'}},
  Phys.~Rev.~D {\bf 83} (2011) 104023,
  [\href{http://arxiv.org/abs/1103.1171}{{\tt arXiv:1103.1171}}].

\bibitem{Cai:1999xg}
R.-G. Cai and N.~Ohta, {\it {`Surface counterterms and boundary stress energy
  tensors for asymptotically nonAnti-de Sitter spaces]}},  Phys.~Rev.~D {\bf
  62} (2000) 024006, [\href{http://arxiv.org/abs/hep-th/9912013}{{\tt
  hep-th/9912013}}].

\bibitem{Levin:2008mq}
J.~Levin and G.~Perez-Giz, {\it {`A Periodic Table for Black Hole Orbits'}},
  Phys.~Rev.~D {\bf 77} (2008) 103005,
  [\href{http://arxiv.org/abs/0802.0459}{{\tt arXiv:0802.0459}}].

\bibitem{Hod:2017xkz}
S.~Hod, {\it {`Upper bound on the radii of black-hole photonspheres'}},
  Phys.~Lett.~B {\bf 727} (2013) 345--348,
  [\href{http://arxiv.org/abs/1701.06587}{{\tt arXiv:1701.06587}}].

\bibitem{Rindler:2007zz}
W.~Rindler and M.~Ishak, {\it {`Contribution of the cosmological constant to
  the relativistic bending of light revisited'}},  Phys.~Rev.~D {\bf 76} (2007)
  043006, [\href{http://arxiv.org/abs/0709.2948}{{\tt arXiv:0709.2948}}].

\end{thebibliography}\endgroup

\end{document}